\documentclass[12pt]{article}

\usepackage[T1]{fontenc}
\usepackage[numbers]{natbib}
\usepackage{txfonts}
\newtheorem{thm}{Theorem}
\newtheorem{cor}[thm]{Corollary}
\newtheorem{lem}[thm]{Lemma}
\newtheorem{assum}[thm]{Assumption}
\newtheorem{prop}[thm]{Proposition}

\newtheorem{exmp}[thm]{Example}

\newcommand{\qed}{\hspace*{\fill}\rule{1.3ex}{1.3ex}}
\newenvironment{pf}{\noindent\emph{Proof.}}{\qed}

\begin{document}
\hypersetup{pdfstartview={FitBH -32768},pdfauthor={Ryutaroh Matsumoto, Diego Ruano and Olav Geil},pdftitle={Generalization of the Lee-O'Sullivan List Decoding for One-Point AG Codes},pdfkeywords={algebraic geometry code, Gr\"obner basis, list decoding}}
\sloppy
\title{Generalization of the Lee-O'Sullivan List Decoding
for One-Point AG Codes\thanks{%
The proposed algorithm in this paper was published
without any proof of
its correctness in Proc.\  2012 IEEE International Symposium on Information Theory,
Cambridge, MA, USA, July 2012, pp.\ 86--90 \citep{gmr12isit}.
This paper was accepted for publication in 
\href{http://www.journals.elsevier.com/journal-of-symbolic-computation/}{Journal of Symbolic Computation}.
This eprint is so-called Accepted Author Manuscript.}}
\author{Ryutaroh Matsumoto\thanks{Department of Communications and Computer Engineering,
Tokyo Instiutte of Technology, 152-8550 Japan}
\and
Diego Ruano\thanks{Department of Mathematical Sciences, Aalborg University, Denmark} \and
Olav Geil\footnotemark[3]}
\date{March 8, 2013}
\maketitle
\begin{abstract}
We generalize the list decoding algorithm for
Hermitian codes proposed by
\citet{lee09} based on Gr\"obner bases
to general one-point AG codes,
under an assumption weaker than one used by \citet{beelen10}.
Our generalization enables us to apply the fast algorithm
to compute a Gr\"obner basis of a module proposed by
\citet{lee09},
which was not possible in another generalization by \citet{lax12}.
\end{abstract}

\section{Introduction}
We consider the list decoding problem of one-point algebraic geometry (AG) codes.
\citet{guruswami99} proposed the well-known
list decoding
algorithm for one-point AG codes, which consists of
the interpolation step and the factorization step.
The interpolation step has large computational complexity and
many researchers have proposed faster interpolation steps,
see \citep[Figure 1]{beelen10}.
\citet{lee09} proposed a faster interpolation step based
on the Gr\"obner basis theory for one-point Hermitian codes.
\citet{beelen10} proposed
the fastest interpolation procedure
for the so-called $C_{ab}$ curves \citep{miura92}
with an additional assumption \citep[Assumptions 1 and 2]{beelen10}.
\citet{little11} generalized the method
in \citet{lee09} to codes defined using a curve
satisfying the same assumption as \citet[Assumptions 1 and 2]{beelen10}.
\citet{lax12} generalized part of \citep{lee09},
namely the interpolation ideal,
to general algebraic curves,
but he did not generalize the faster interpolation algorithm in \citep{lee09}.
The aim of this paper is to generalize the
faster interpolation algorithm
\citep{lee09} to an even wider class of algebraic curves
than  \citep{little11}.
%We shall demonstrate that our proposal
%provides much faster alternative to
We shall compare our proposal with
the previously known interpolation algorithms for
the code on the Klein quartic in Example \ref{ex4}.
As a byproduct of our argument,
in Corollary \ref{cor1}
we also clarifies the relation between two different definitions
of modules used by \citet{sakata01} and
by \citet{lax12,lee09} for list decoding.

This paper is organized as follows:
Section \ref{sec2} introduces notations and relevant facts.
Section \ref{sec3} generalizes \citep{lee09}.
Section \ref{sec4} concludes the paper.

\section{Notation and Preliminary}\label{sec2}
Our study heavily relies on the standard form of algebraic curves
introduced independently by \citet{geilpellikaan00} and 
\citet{miura98},
which is an enhancement of earlier results \citep{miura92,saints95}.
Let $F/\mathbf{F}_q$ be an algebraic function field of one variable over
a finite field $\mathbf{F}_q$ with $q$ elements.
Let $g$ be the genus of $F$.
Fix $n+1$ distinct places $Q$, $P_1$, \ldots,
$P_n$ of degree one in $F$ and a nonnegative integer
$u$. We consider the following
one-point algebraic geometry (AG) code
\[
C_u = \{ (f(P_1), \ldots, f(P_n)) \mid f \in \mathcal{L}(uQ)\}.
\]
Suppose that the Weierstrass semigroup $H(Q)$
at $Q$ is generated by
$a_1$, \ldots, $a_t$, and choose
$t$ elements $x_1$, \ldots,
$x_t$ in $F$ whose pole divisors are $(x_i)_\infty = a_iQ$ for $i=1$, \ldots, $t$.
%We do \emph{not} assume that
%$a_1$ is the smallest among $a_1$, \ldots, $a_t$.
Without loss of generality we may assume the availability of
such $x_1$, \ldots,
$x_t$, because otherwise we cannot find a basis of $C_u$ for every $u$,
i.e.\ we cannot construct the code $C_u$.
Then we have that $\mathcal{L}(\infty Q) = \cup_{i=1}^\infty\mathcal{L}(iQ)$
is equal to $\mathbf{F}_q[x_1$, \ldots, $x_t]$ \citep{saints95}.
We express $\mathcal{L}(\infty Q)$ as a residue class ring
$\mathbf{F}_q[X_1$, \ldots, $X_t]/I$
of the polynomial ring $\mathbf{F}_q[X_1$, \ldots, $X_t]$, where
$X_1$, \ldots, $X_t$ are transcendental over $\mathbf{F}_q$,
and $I$ is the kernel of the canonical homomorphism sending
$X_i$ to $x_i$. \citet{geilpellikaan00} and \citet{miura98}
identified the following convenient representation of
$\mathcal{L}(\infty Q)$ by using the Gr\"obner basis theory \citep{adams94}.
The following review is borrowed from \citep{miuraform}.
Hereafter, we assume that the reader is familiar with the
Gr\"obner basis theory in \citep{adams94}.

Let $\mathbf{N}_0$ be the set of nonnegative integers.
For $(m_1$, \ldots, $m_t)$, $(n_1$, \ldots, $n_t) \in
\mathbf{N}_0^t$,
we define the weighted reverse lexicographic monomial order $\succ$
such that $(m_1$, \ldots, $m_t)$ $\succ$ $(n_1$, \ldots, $n_t)$
if $a_1 m_1 + \cdots + a_t m_t > a_1 n_1 + \cdots + a_t n_t$,
or $a_1 m_1 + \cdots + a_t m_t = a_1 n_1 + \cdots + a_t n_t$,
and $m_1 = n_1$, $m_2 = n_2$, \ldots,
$m_{i-1} = n_{i-1}$, $m_i<n_i$, for some $1 \leq i \leq t$.
Note that a Gr\"obner basis of $I$ with respect to $\succ$
can be computed by
\citep[Theorem 15]{saints95}, \citep{schicho98},
\citep[Theorem 4.1]{tang98} or
\citep[Proposition 2.17]{bn:vasconcelos},
starting from any affine
defining equations of $F/\mathbf{F}_q$.

\begin{exmp}\label{ex1}
According to \citet[Example 3.7]{hoholdt95},
\[
u^3 v + v^3 + u = 0
\]
is an affine defining equation for the Klein quartic over
$\mathbf{F}_8$.
There exists a unique $\mathbf{F}_8$-rational place $Q$ such that
$(v)_\infty= 3Q$, $(uv)_\infty = 5Q$, and $(u^2v)_\infty
= 7Q$. The numbers $3$, $5$ and $7$ constitute the minimal
generating set of the Weierstrass semigroup at $Q$.
Choosing $x_1$ as $v$, $x_2$ as $uv$ and $x_3$ as $u^2v$,
by \citet[Theorem 4.1]{tang98} we can see that the standard form of the Klein quartic is
given by
\[
X_2^2+X_3X_1,\quad X_3X_2+X_1^4+X_2,\quad X_3^2+X_2X_1^3+X_3,
\]
which is the reduced Gr\"obner basis for $I$
with respect to the monomial order $\succ$.
We can see that $a_1=3$, $a_2=5$, and $a_3=7$.
\end{exmp}

For $i=0$, \ldots, $a_1-1$,
we define $b_i = \min\{ m \in H(Q) \mid m \equiv i \pmod{a_1}\}$,
and $L_i$ to be the minimum element $(m_1$, \ldots,
$m_t) \in \mathbf{N}_0^t$
with respect to $\prec$ such that $a_1 m_1 + \cdots +
a_t m_t = b_i$.
Note that the set of $b_i$'s is the well-known Ap\'ery set \citep{MR0017942} and \citep[Lemmas 2.4 and 2.6]{MR2549780} of the numerical semigroup $H(Q)$.
Then we have $\ell_1 = 0$ if we write
$L_i$ as $(\ell_1$, \ldots, $\ell_t)$.
For each $L_i = (0$, $\ell_{i2}$, \ldots, $\ell_{it})$,
define $y_i = x_2^{\ell_{i2}} \cdots x_t^{\ell_{it}} \in \mathcal{L}(\infty Q)$.

The footprint of $I$, denoted by $\Delta(I)$,
is $\{ (m_1$, \ldots, $m_t) \in \mathbf{N}_0^t \mid X_1^{m_1} \cdots
X_t^{m_t}$ is not the leading monomial of
any nonzero polynomial in $I$ with respect to $\prec\}$,
and define $B = \{x_1^{m_1} \cdots x_t^{m_t} \mid 
(m_1$, \ldots, $m_t) \in \Delta(I)\}$.
Then $B$ is a basis of $\mathcal{L}(\infty Q)$
as an  $\mathbf{F}_q$-linear space \citep{adams94},
two distinct elements in $B$ have
different pole orders at $Q$,
and
\begin{eqnarray}
B &=& \{ x_1^m x_2^{\ell_2} \cdots, x_t^{\ell_t} \mid m \in \mathbf{N}_0,
(0, \ell_2, \ldots, \ell_t) \in \{L_0, \ldots, L_{a_1-1}\}\}\nonumber\\
&=& \{ x_1^m y_i \mid m \in \mathbf{N}_0, i=0, \ldots, a_1-1\}.
\label{eq:footprintform}
\end{eqnarray}
Equation (\ref{eq:footprintform}) shows that
$\mathcal{L}(\infty Q)$ is a free $\mathbf{F}_q[x_1]$-module
with a basis $\{y_0$, \ldots, $y_{a_1-1}\}$.
Note that the above structured shape of $B$ reflects the well-known
property of every weighted reverse lexicographic monomial order,
see the paragraph preceding to \citep[Proposition 15.12]{bn:eisenbud}.
\begin{exmp}\label{ex2}
For the curve in Example \ref{ex1},
we have $y_0 = 1$, $y_1=x_3$, $y_2=x_2$.
\end{exmp}

Let $v_Q$ be the unique valuation in $F$ associated with the place $Q$.
The semigroup $H(Q)$ is equal to $\{i a_1 - v_Q (y_j) \mid 0\le i,0\le j<a_1\}$ \citep[Lemma 2.6]{MR2549780}.

\section{Generalization of Lee-O'Sullivan's List Decoding to General One-Point AG Codes}\label{sec3}
\subsection{Background on Lee-O'Sullivan's Algorithm}
In the famous list decoding algorithm for the one-point AG
codes in \citep{guruswami99},
we have to compute the univariate interpolation polynomial
whose coefficients belong to $\mathcal{L}(\infty Q)$.
\citet{lee09}
proposed a faster algorithm to compute
the interpolation polynomial for the Hermitian one-point codes.
Their algorithm was sped up and generalized to one-point AG codes
over the so-called $C_{ab}$ curves \citep{miura92} by \citet{beelen10}
with an additional assumption.
In this section we generalize Lee-O'Sullivan's procedure to
general one-point AG codes with an  assumption weaker than
\citep[Assumption 2]{beelen10},
which will be introduced in and used after Assumption \ref{assump1}.
The argument before Assumption \ref{assump1} is true
without Assumption \ref{assump1}.

Let $m$ be the multiplicity parameter in \citep{guruswami99}.
\citet{lee09} introduced the ideal $I_{\vec{r},m}$ for Hermitian curves
containing the interpolation polynomial corresponding to
the received word $\vec{r}$ and the multiplicity $m$.
The ideal $I_{\vec{r},m}$ contains the interpolation polynomial
as its  nonzero element minimal with respect to the weighted reverse lexicographic monomial order $\prec_u$ to be
introduced in
Section \ref{sec32}. 
We will give a generalization of $I_{\vec{r},m}$ for general
algebraic curves.

\subsection{Generalization of the Interpolation Ideal}
Let $\vec{r} = (r_1$, \ldots, $r_n) \in \mathbf{F}_q^n$
be the  received word.
For a divisor $G$ of $F$, we define
$\mathcal{L}(-G + \infty Q) = \bigcup_{i=1}^\infty \mathcal{L}(-G+iQ)$.
We see that $\mathcal{L}(-G + \infty Q)$ is an ideal of $\mathcal{L}(\infty Q)$
\citep{matsumoto99ldpaper}.

Let $h_{\vec{r}} \in \mathcal{L}(\infty Q)$ such that $h_{\vec{r}}(P_i) = r_i$.
Computation of such $h_{\vec{r}}$ can be easily done as
follows provided that
we can construct generator matrices for $C_u$
for all $u$.
For $1 \leq j \leq n$, define $\psi_j \in B$ such that
$\dim C_{-v_Q(\psi_j)} = j$, and let
\[
\left(\begin{array}{c}
i_1\\
\vdots\\
i_n\end{array}\right)
=
\left(
\begin{array}{ccc}
\psi_1(P_1)&\cdots&\psi_1(P_n)\\
\vdots&\vdots& \vdots\\
\psi_n(P_1)&\cdots&\psi_n(P_n)
\end{array}\right)^{-1} \vec{r}.
\]
We find that $h_{\vec{r}} = \sum_{j=1}^n i_j \psi_j$
satisfies the required condition for $h_{\vec{r}}$.
Since $-v_Q(\psi_n) \leq n+2g-1$,
we can choose $h_{\vec{r}}$ so that $-v_Q(h_{\vec{r}}) \leq n+ 2g-1$.

Let $Z$ be transcendental over $\mathcal{L}(\infty Q)$,
and $D=P_1 + \cdots + P_n$.
$\mathcal{L}(\infty Q)[Z]$ denotes
the univariate polynomial ring of $Z$ over $\mathcal{L}(\infty Q)$.
For a divisor $G$
we denote by $\mathcal{L}_Z(-G +\infty Q)$ the ideal
of $\mathcal{L}(\infty Q)[Z]$ generated by
$\mathcal{L}(-G +\infty Q) \subset \mathcal{L}(\infty Q)$.
Define the ideal $I_{\vec{r},m}$ of $\mathcal{L}(\infty Q)[Z]$ as
\begin{eqnarray}
I_{\vec{r},m} &=& \mathcal{L}_Z(-mD +\infty Q) + \mathcal{L}_Z(-(m-1)D +\infty Q)
\langle Z-h_{\vec{r}} \rangle + \cdots \nonumber\\
&& + \mathcal{L}_Z(-D +\infty Q) \langle Z-h_{\vec{r}} \rangle^{m-1} + \langle Z-h_{\vec{r}} \rangle^{m}, \label{eq20}
\end{eqnarray}
where $\langle \cdot \rangle$ denotes the ideal generated by $\cdot$,
the plus sign $+$ denotes the sum of ideals,
and $\mathcal{L}_Z(-iD +\infty Q) \langle Z-h_{\vec{r}} \rangle^{m-i}$ denotes the product of two ideals
$\mathcal{L}_Z(-iD +\infty Q)$ and $\langle Z-h_{\vec{r}} \rangle^{m-i}$.
We remark that the above $I_{\vec{r},m}$ is equal
to $\bar{I}_{m,v}$
defined by \citet{lax12}.
Note that our definition does not involve coordinate
variables $x_1$, $x_2$, \ldots of the defining equations
as used by \citet{lax12}.
For $Q(Z) \in \mathcal{L}(\infty Q)[Z]$, we say $Q(Z)$ has multiplicity $m$ at $(P_i, r_i)$
if 
\begin{equation}\label{eq21}
Q(Z+r_i) = \sum_{j} \alpha_j Z^j 
\end{equation}
with $\alpha_j \in \mathcal{L}(\infty Q)$
satisfies $v_{P_i}(\alpha_j) \geq m-j$ for all $j$.
\citet[Section 3.2]{sakata01} introduced
a special case of the following set
for Hermitian curves. We give a more general definition
(for any curve) as follows:
\[
I'_{\vec{r},m} = \{ Q(Z) \in \mathcal{L}(\infty Q)[Z] \mid Q(Z)\mbox{ has multiplicity $m$ for
all }(P_i, r_i) \}.
\]
This definition of the multiplicity is the same as \citep{guruswami99}.
Therefore, we can find the interpolation polynomial used in \citep{guruswami99}
from $I'_{\vec{r},m}$.
We shall explain how to find efficiently the interpolation polynomial
from $I'_{\vec{r},m}$, after clarifying the relation between
$I_{\vec{r},m}$ and $I'_{\vec{r},m}$.

\begin{lem}\label{lem2}
We have $I_{\vec{r},m} \subseteq I'_{\vec{r},m}$.
\end{lem}
\begin{pf}
Observe that $I'_{\vec{r},m}$ is an ideal of $\mathcal{L}(\infty Q)[Z]$.
Let $\alpha (Z-h_{\vec{r}})^j \in \mathcal{L}_Z(-(m-j)D +\infty Q)
\langle Z-h_{\vec{r}} \rangle^j$ such that $\alpha \in \mathcal{L}(-(m-j)D +\infty Q)$.
Then we have
\[
\alpha (Z+r_i - h_{\vec{r}})^j
= \alpha (Z-(h_{\vec{r}}-r_i))^j\\
= \sum_{k=0}^j \alpha_k (h_{\vec{r}}-r_i)^{j-k} Z^k ,
\]
where $\alpha_k \in \mathcal{L}(-(m-j)D +\infty Q)$.
We can see that $\alpha_k (h_{\vec{r}}-r_i)^{j-k} \in \mathcal{L}(-(m-k)P_i +\infty Q)$
and that $\mathcal{L}(-(m-j)D +\infty Q)
\langle Z-h_{\vec{r}} \rangle^j \subseteq I'_{\vec{r},m}$,
because
$\mathcal{L}_Z(-(m-j)D +\infty Q) \langle Z-h_{\vec{r}} \rangle^{j}$
is generated by $\{ \alpha (Z-h_{\vec{r}})^{j} \mid \alpha \in \mathcal{L}(-(m-j)D +\infty Q)\}$ as an ideal of $\mathcal{L}(\infty Q)[Z]$.
Since $I'_{\vec{r},m}$ is an ideal, it follows that
$I_{\vec{r},m} \subseteq I'_{\vec{r},m}$.
\end{pf}

The following Proposition \ref{prop1} will be used in the proof of
Proposition \ref{prop2}.
\begin{prop}\label{prop1}\citep{guruswami99}
$\dim_{\mathbf{F}_q} \mathcal{L}(\infty Q)[Z]/I'_{\vec{r},m} = n {m+1 \choose 2}$.
\end{prop}

\begin{lem}\label{lem1}
Let $G$ be a divisor $\geq 0$ whose support is disjoint from $Q$.
If $\deg P = 1$ for all $P \in \mathrm{supp}(G)$ then
we have
\[
\dim_{\mathbf{F}_q} \mathcal{L}(\infty Q) / \mathcal{L}(-G + \infty Q) = \deg G.
\]
\end{lem}
\begin{pf}
Let $n()$ be a mapping from $\mathrm{supp}(G)$ to the set of 
nonnegative integers.
Let $\mathcal{N}$ be the set of those functions such that
$n(P) < v_P (G)$ for all $P\in \mathrm{supp}(G)$.
By the strong approximation theorem \citep[Theorem I.6.4]{bn:stichtenoth}
we can choose a $f_{n()} \in
\mathcal{L}(\infty Q)$ such that $v_P(f_{n()}) = n(P)$ for
every $P \in \mathrm{supp}(G)$.
Any element in $\mathcal{L}(\infty Q) \setminus \mathcal{L}(-G + \infty Q)$
can be written as the sum of an element $g \in  \mathcal{L}(-G + \infty Q)$
plus an $\mathbf{F}_q$-linear
combination of $f_{n()}$'s
by the assumption $\deg P = 1$ for all $P \in \mathrm{supp}(G)$,
which completes the proof.
\end{pf}

The following proposition is equivalent to
\citet[Proposition 6]{lax12}, but we include its proof
because our definition of $I_{\vec{r},m}$
is apparently very different from that of $\bar{I}_{m,v}$ by
\citet{lax12}.
\begin{prop}\label{prop2}
$\dim_{\mathbf{F}_q} \mathcal{L}(\infty Q)[Z]/I_{\vec{r},m} = n {m+1 \choose 2}$.
\end{prop}
\begin{pf}
Recall that  $I$ is an ideal of $\mathbf{F}_q[X_1$, \ldots, $X_t]$
such that $\mathcal{L}(\infty Q) = \mathbf{F}_q[X_1$, \ldots, $X_t]/I$
as introduced in  Section \ref{sec2}.
Let $G_i$ be a Gr\"obner basis of the preimage of $\mathcal{L}(-iD +\infty Q)$
in $\mathbf{F}_q[X_1$, \ldots, $X_t]$,
and $H_{\vec{r}}$ be the coset representative of $h_{\vec{r}}$
written as a sum of monomials whose exponents belong to $\Delta(I)$.
In this proof,
the footprint $\Delta(\cdot)$ is always considered for
$\mathbf{F}_q[X_1$, \ldots, $X_t]$ excluding the variable $Z$.
Then
\[
G = \cup_{i=0}^m \{ F (Z - H_{\vec{r}})^{m-i} \mid F \in G_i \}
\]
is a 
 Gr\"obner basis of the preimage of $I_{\vec{r},m}$
in $\mathbf{F}_q[Z, X_1$, \ldots, $X_t]$ with the elimination
monomial order with $Z$ greater than $X_i$'s and refining the monomial
order $\succ$ defined in  Section \ref{sec2}.
Please refer to \citep[Section 15.2]{bn:eisenbud} for refining monomial
orders.
A remainder of division by $G$ can always be written as
\[
F_{m-1} Z^{m-1} + F_{m-2}Z^{m-2} + \cdots + F_0
\]
with $F_i \in \mathbf{F}_q[X_1$, \ldots, $X_t]$.
Then $F_{m-i}$ must be written as
a sum of monomials
whose exponents belong to the footprint $\Delta (G_i)$ of
$G_i$, for $i=1$, \ldots, $m$.
This shows that
\[
\dim_{\mathbf{F}_q} \mathcal{L}(\infty Q)[Z] / I_{\vec{r},m} \leq
\sum_{i=1}^{m} \sharp \Delta(G_i).
\]
On the other hand, by Lemma \ref{lem1},
\[
\sharp \Delta(G_i) = \dim_{\mathbf{F}_q} \mathcal{L}(\infty Q) / \mathcal{L}(-iD + \infty Q) = ni.
\]
This implies
\[
\dim_{\mathbf{F}_q} \mathcal{L}(\infty Q)[Z] / I_{\vec{r},m} \leq n {m+1 \choose 2}.
\]
By Proposition \ref{prop1} and Lemma \ref{lem2}, we
see 
\[
\dim_{\mathbf{F}_q} \mathcal{L}(\infty Q)[Z] / I_{\vec{r},m} = n {m+1 \choose 2}.
\]
\end{pf}

The following corollary clarifies the relation between
the module $I'_{\vec{r},m}$ used by \citet{sakata01}
and $I_{\vec{r},m}$ used by \citet{lax12,lee09},
which was not explicit in previous literature.
\begin{cor}\label{cor1}
$I'_{\vec{r},m} = I_{\vec{r},m}$. \qed
\end{cor}
Since $I'_{\vec{r},m}$ is the ideal used in \citep{guruswami99},
we can find the required interpolation polynomial
directly from an $\mathbf{F}_q[x_1]$-submodule of $I_{\vec{r},m}=I'_{\vec{r},m}$
as explained in Section \ref{sec32}.

For $i=0$, \ldots, $m$ and $j=0$, \ldots, $a_1-1$,
let $\eta_{i,j}$ to be an element in $\mathcal{L}(-iD+\infty Q)$
such that $-v_Q(\eta_{i,j})$ is the minimum among $\{
-v_Q(\eta) \mid \eta \in \mathcal{L}(-iD+\infty Q)$,
$ -v_Q(\eta) \equiv j \pmod{a_1}\}$.
Such elements $\eta_{i,j}$ can be computed by \citep{matsumoto99ldpaper}
before receiving $\vec{r}$.
It was also shown \citep{matsumoto99ldpaper} that
$\{ \eta_{i,j} \mid j=0$, \ldots, $a_1-1 \}$ generates
$\mathcal{L}(-iD+\infty Q)$ as an $\mathbf{F}_q[x_1]$-module.
Note also that we can choose $\eta_{0,i} = y_i$ defined in Section \ref{sec2}.
By Eq.\ (\ref{eq:footprintform}),
all $\eta_{i,j}$
and $h_{\vec{r}}$ can be
expressed as polynomials in $x_1$ and $y_0$, \ldots, $y_{a_1-1}$.
Thus we have
\begin{thm}[{Generalization of \citet[Proposition 6]{beelen10} and \citet{little11}}]\label{thm:generators}
Let $\ell \geq m$.
One has that
\begin{eqnarray*}
&&\{ (Z-h_{\vec{r}})^{m-i}\eta_{i,j} \mid i=0, \ldots, m, j=0, \ldots, a_1-1 \}\\
&\cup& \{ Z^{\ell-m}(Z-h_{\vec{r}})^{m}\eta_{0,j} \mid \ell=1, \ldots, j=0, \ldots, a_1-1 \}
\end{eqnarray*}
generates
\[
I_{\vec{r},m,\ell} = I_{\vec{r},m} \cap \{ Q(Z) \in \mathcal{L}(\infty Q)[Z] \mid \deg_Z Q(Z) \leq \ell \}
\]
as an $\mathbf{F}_q[x_1]$-module.
\end{thm}
\begin{pf}
Let $e \in I_{\vec{r},m}$ and $E$ be its preimage in
$\mathbf{F}_q[Z$, $X_1$, \ldots, $X_t]$.
By dividing $E$ by the Gr\"obner basis $G$ introduced in
the proof of Proposition \ref{prop2}, we can see that
$e$ is expressed as
\[
e= \sum_{\ell=1} \alpha_{-\ell} Z^\ell (Z-h_{\vec{r}})^m + \sum_{i=0}^m \alpha_i(Z-h_{\vec{r}})^{m-i}
\]
with $\alpha_i \in \mathcal{L}(-\max\{i,0\}D + \infty Q)$,
from which the assertion follows.
\end{pf}

\subsection{Computation of the Interpolated Polynomial from the
Interpolation Ideal $I_{\vec{r},m}$}\label{sec32}
For $(m_1$, \ldots, $m_t$, $m_{t+1})$, $(n_1$, \ldots, $n_t$, $n_{t+1}) \in
\mathbf{N}_0^{t+1}$,
we define the other weighted reverse lexicographic monomial order $\succ_u$
in $\mathbf{F}_q[X_1$, \ldots, $X_t$, $Z]$
such that $(m_1$, \ldots, $m_t$, $m_{t+1})$ $\succ_u$ $(n_1$, \ldots, $n_t$, $n_{t+1})$
if $a_1 m_1 + \cdots + a_t m_t + u m_{t+1}> a_1 n_1 + \cdots + a_t n_t + un_{t+1}$,
or $a_1 m_1 + \cdots + a_t m_t + u m_{t+1} = a_1 n_1 + \cdots + a_t n_t + u n_{t+1}$,
and $m_1 = n_1$, $m_2 = n_2$, \ldots,
$m_{i-1} = n_{i-1}$, $m_i<n_i$, for some $1 \leq i \leq t+1$.
As done in \citep{lee09},
the interpolation polynomial is the smallest nonzero polynomial
with respect to $\succ_u$ in the preimage of $I_{\vec{r},m}$.
Such a smallest element can be found from a Gr\"obner basis of the
$\mathbf{F}_q[x_1]$-module $I_{\vec{r},m,\ell}$
in Theorem \ref{thm:generators}.
To find such a Gr\"obner basis, Lee and O'Sullivan proposed the following
general purpose algorithm as \citep[Algorithm G]{lee09}.

Their algorithm \citep[Algorithm G]{lee09}
efficiently finds a Gr\"obner basis of submodules of $\mathbf{F}_q[x_1]^s$
for a special kind of generating set and monomial orders.
Please refer to \citep{adams94} for Gr\"obner bases for modules.
Let $\mathbf{e}_1$, \ldots, $\mathbf{e}_s$ be the standard basis
of $\mathbf{F}_q[x_1]^s$.
Let $u_x$, $u_1$, \ldots, $u_s$ be positive integers.
Define the monomial order in the $\mathbf{F}_q[x_1]$-module
$\mathbf{F}_q[x_1]^s$
such that $x_1^{n_1} \mathbf{e}_i \succ_{\mathrm{LO}} x_1^{n_2} \mathbf{e}_j$
if ${n_1}u_x + u_i > {n_2}u_x + u_j$ or ${n_1}u_x + u_i = {n_2}u_x + u_j$
and $i>j$.
For $f = \sum_{i=1}^s f_i(x_1)\mathbf{e}_i \in \mathbf{F}_q[x_1]^s$,
define $\mathrm{ind}(f) = \max\{i \mid f_i(x_1) \neq 0\}$,
where $f_i(x_1)$ denotes a univariate polynomial in $x_1$ over $\mathbf{F}_q$.
Their algorithm \citep[Algorithm G]{lee09}
efficiently computes a Gr\"obner basis
with respect to $\succ_{\mathrm{LO}}$ of a module generated
by $g_1$, \ldots, $g_s \in \mathbf{F}_q[x_1]^s$ such that
$\mathrm{ind}(g_i) = i$.
The computational complexity is also evaluated in \citep[Proposition 16]{lee09}.

Let $\ell$ be the maximum $Z$-degree of the interpolation polynomial
in \citep{guruswami99}. 
The set $I_{\vec{r},m,\ell}$ in Theorem \ref{thm:generators}
is an $\mathbf{F}_q[x_1]$-submodule of $\mathbf{F}_q[x_1]^{a_1(\ell+1)}$
with the module basis $\{ y_j Z^k \mid j=0$, \ldots, $a_1-1$, $k=0$, \ldots,
$\ell \}$.

\begin{assum}\label{assump1}
We assume that there exists $f \in \mathcal{L}(\infty Q)$ whose
zero divisor $(f)_0 = D$.
\end{assum}
By the algorithm of \citet{matsumoto99ldpaper},
we can find $f$ in Assumption \ref{assump1} if it exists.

The assumptions in \citep{beelen10}
are
\begin{itemize}
\item The function field $F$ was defined by a nonsingular
affine algebraic curve of the form
\begin{equation}
\gamma_{a_2,0} X_1^{a_2} + \gamma_{0,a_1} X_2^{a_1}
+ \sum_{i a_2 + j a_1 < a_1 a_2} \gamma_{i,j} X_1^i X_2^j
\label{eq:cab}
\end{equation}
with $\mathrm{gcd}(a_1$, $a_2)=1$, $\gamma_{a_2,0} \neq 0$ and
$\gamma_{0,a_1} \neq 0$,
\item and Assumption \ref{assump1} above.
\end{itemize}
Since the function field can be defined in the form
(\ref{eq:cab}) if the Weierstrass semigroup $H(Q)$
is generated by relatively prime positive integers $a_1$ and $a_2$
\citep{miuraform},
we can see that Assumption \ref{assump1} is implied by \citep[Assumption 2]{beelen10} 
and is weaker than \citep[Assumption 2]{beelen10}.

Let $\langle f \rangle$ be the ideal of $\mathcal{L}(\infty Q)$
generated by $f$. By \citep[Corollary 2.3]{matsumoto99ldpaper}
we have $\mathcal{L}(-D + \infty Q) = \langle f \rangle$.
By \citep[Corollary 2.5]{matsumoto99ldpaper}
we have $\mathcal{L}(-iD + \infty Q) = \langle f^i \rangle$.

\begin{exmp}\label{ex3}
This is continuation of Example \ref{ex2}.
Let $f=x_1^7+1$. We see that $-v_Q(f) = 21$ and that
there exist $21$ distinct $\mathbf{F}_8$-rational
places $P_1$, \ldots, $P_{21}$, such that $f(P_i)=0$
for $i=1$, \ldots, $21$ by straightforward computation.
By setting $D=P_1 + \cdots + P_{21}$ Assumption \ref{assump1} is
satisfied.

We remark that we have $-v_Q(x_1^8+x_1)=24$ but there exist only
$23$ $\mathbf{F}_8$-rational places $P$ such that
$(x_1^8+x_1)(P)=0$, other than $Q$, and that $(x_1^8+x_1)$
does not satisfy Assumption \ref{assump1}.
\end{exmp}

Without loss of generality we may assume existence of
$x' \in \mathcal{L}(\infty Q)$ such that $f \in \mathbf{F}_q[x']$,
because we can set $x'=f$.
By changing the choice of $x_1$, \ldots, $x_t$ if necessary,
we may assume $x_1 = x'$ and $f \in \mathbf{F}_q[x_1]$ without
loss of generality, while it is better to make $-v_Q(x_1)$ as small
as possible in order to reduce the computational complexity.
Under the assumption $f \in \mathbf{F}_q[x_1]$,
$f^i y_j$ satisfies the required condition for
$\eta_{i,j}$ in Theorem \ref{thm:generators}.
By naming $y_j Z^k$ as $\mathbf{e}_{1+j+ku}$,
the generators in Theorem \ref{thm:generators} satisfy the
assumption in \citep[Algorithm G]{lee09}.
In the following, we assign weight $-iv_Q(x_1)-v_Q(y_j)+ku$
to the module element
$x_1^iy_jZ^k$. With this assignment of weights,
the monomial order $\succ_{\mathrm{LO}}$
is the restriction of $\succ_u$ to the $\mathbf{F}_q[x_1]$-submodule
of $\mathcal{L}(\infty Q)[Z]$ generated by $\{ y_j Z^k \mid
j=0$, \ldots, $a_1-1$, $k=0$, \ldots, $\ell\}$.
We can efficiently compute a Gr\"obner basis of
the $\mathbf{F}_q[x_1]$-module $I_{\vec{r},m,\ell}$
by \citep[Algorithm G]{lee09}.
After that we find
the interpolation polynomial required in the list decoding algorithm
by \citet{guruswami99} as the minimal
element with respect to $\succ_{\mathrm{LO}}$
in the computed Gr\"obner basis.

\begin{prop}
Suppose that
we use \cite[Algorithm G]{lee09} to find
the Gr\"obner basis of $I_{\vec{r},m,\ell}$ with respect to
$\succ_{\mathrm{LO}}$.
Under Assumption \ref{assump1},
the number of multiplications in \cite[Algorithm G]{lee09}
with the generators in Theorem \ref{thm:generators} 
is at most 
\begin{equation}
[\max_j\{-v_Q(y_j)\} + m (n+2g-1)+ u(\ell-m)]^2a_1^{-1}
\sum_{i=1}^{a_1(\ell + 1)} i^2. \label{eq:compl}
\end{equation}
\end{prop}
\begin{pf}
What we shall do in this proof
is substitution of variables in the general complexity
formula in \citet{lee09} by specific values.
The number of generators is $a_1(\ell + 1)$, which is denoted by
$m$ in  \cite[Proposition 16]{lee09}.
We have $-v_Q(f) \leq n+g$ and $-v_Q(h_{\vec{r}}) \leq n+2g-1$.
We can assume $u\leq n+2g-1$.
Thus, the maximum weight of the generators is upper bounded by
\[
\max_j\{-v_Q(y_j)\} + m (n+2g-1)+ u(\ell-m).
\]
By \cite[Proof of Proposition 16]{lee09}, the number of multiplications
is upper bounded by Eq.\ (\ref{eq:compl}).
\end{pf}

\begin{exmp}\label{ex4}
Consider the $[21,10]$
code $C_{12}$ over the Klein quartic considered in
Examples \ref{ex1}, \ref{ex2} and \ref{ex3}.
Its Goppa bound is $n-u = 21-12=9$.
The equivalent algorithms by \citet{beelen08,guruswami99}
can correct $5$
errors with $m=40$ and $\ell=54$.
An advantage of
\citet{beelen08} over \citet{guruswami99} is that
the former solves a smaller system of linear equations
by utilizing the structure of the equations,
and thus is faster than the latter.

We shall evaluate the number of multiplications and divisions
by the method in \cite{beelen08}.
One can choose the divisor $A$ in \citep[Section 2.6]{beelen08}
as $(m(n-5)-1)Q = 639Q$.
The algorithm by \citet{beelen08} solves a system of
\begin{eqnarray*}
&& \sum_{i=0}^m((m-i)n - \dim (A-iu Q)+ \dim (-(m-i)D + A - iu Q))\\
&=& \sum_{i=0}^{40}(21(40-i) - \dim (639-12i) Q+ \dim (-(40-i)D + (639-12i) Q)\\
&=& 2392
\end{eqnarray*}
linear equations with
\begin{eqnarray*}
&& \sum_{i=m+1}^\ell \dim (A-iu Q) + \sum_{i=0}^m \dim (-(m-i)D + A - iu Q))\\
&=& \sum_{i=41}^{54} \dim (639-12i) Q +
\sum_{i=0}^{40} \dim (-(40-i)D + (639-12i) Q)\\
&=& 2399
\end{eqnarray*}
unknowns. The number of multiplications and divisions is
about $2399^3/3 \simeq 4.6 \times 10^9$.

On the other hand,
The original algorithm 
by \citet{guruswami99}
requires us to solve a system of
$21 \times {40 + 1 \choose 2} =
17220$ linear equations.
Solving such a system needs roughly $17220^3/3 \simeq 1.7 \times 10^{12}$
multiplications and divisions
in $\mathbf{F}_8$.

The value of  Eq.\ (\ref{eq:compl}) is given by
\begin{eqnarray*}
&&[\max_j\{-v_Q(y_j)\} + m (n+2g-1)+ u(\ell-m)]^2a_1^{-1}
\sum_{i=1}^{a_1(\ell + 1)} i^2\\
&=& [7 + 40\cdot 26 + 12(54-40)]^2 / 3 
\times \sum_{i=1}^{3\cdot 55}i^2\\
&=& 28,038,433,500 \simeq 2.8 \times 10^{10}.
\end{eqnarray*}
We see that the proposed method can solve the interpolation step
faster than \citet{guruswami99}, but the method by
\citet{beelen08} is even faster.
\end{exmp}

\section{Concluding Remarks}\label{sec4}
The interpolation step in \citet{guruswami99}
is computationally costly and many researchers proposed
faster interpolation methods, as summarized by
\citet[Figure 1]{beelen10}.
However, except \citet{beelen08},
those researches assumed either Hermitian curves,
e.g.\ \citet{lee09,sakata01} or $C_{ab}$ curves e.g.\ \cite{beelen10,little11}.
Our argument used no assumption until Assumption \ref{assump1}
that seems indispensable with application of Algorithm G in \citet{lee09}.
The Klein quartic is the well-known family for constructing
AG codes.
In Example \ref{ex4} we demonstrated that the proposed interpolation
procedure is faster than the original \citep{guruswami99}
and comparable to \citep{beelen08}
for codes on the Klein quartic.

\section*{Acknowledgment}
The authors would like to thank an anonymous reviewer
for his/her very careful reading of the
initial manuscript,
and Prof.\ Robert Lax for pointing out its errors.
This research was partly supported by  the MEXT Grant-in-Aid for Scientific Research (A) No.\ 23246071, the
 Villum Foundation through their VELUX Visiting Professor Programme
 2011--2012, the Danish National Research Foundation and the National
 Science Foundation of China (Grant No.\ 11061130539) for the
 Danish-Chinese Center for Applications of Algebraic Geometry in
 Coding Theory and Cryptography and by Spanish grant MTM2007-64704
  and by Spanish MINECO grant
No.\ MTM2012-36917-C03-03


\begin{thebibliography}{23}
\providecommand{\href}[2]{#2}
\newcommand{\doi}[1]{doi: \href{http://dx.doi.org/#1}{\path{#1}}}
\providecommand{\natexlab}[1]{#1}
\providecommand{\url}[1]{\texttt{#1}}

\bibitem[Adams and Loustaunau(1994)]{adams94}
W.~W. Adams and P.~Loustaunau.
\newblock \emph{An Introduction to Gr\"obner Bases}, volume~3 of
  \emph{Graduate Studies in Mathematics}.
\newblock American Mathematical Society, Providence, RI, 1994.

\bibitem[Ap\'ery(1946)]{MR0017942}
R.~Ap\'ery.
\newblock Sur les branches superlin\'eaires des courbes alg\'ebriques.
\newblock \emph{C. R. Acad. Sci. Paris}, 222:\penalty0 1198--1200, 1946.

\bibitem[Beelen and Brander(2010)]{beelen10}
P.~Beelen and K.~Brander.
\newblock Efficient list decoding of a class of algebraic-geometry codes.
\newblock \emph{Adv. Math. Commun.}, 4\penalty0 (4):\penalty0 485--518, 2010.

\bibitem[Beelen and H\o holdt(2008)]{beelen08}
P.~Beelen and T.~H\o holdt.
\newblock The decoding of algebraic geometry codes.
\newblock In E.~Mart\'inez-Moro, C.~Munuera, and D.~Ruano, editors,
  \emph{Advances in Algebraic Geometry Codes}, volume~5 of \emph{Coding Theory
  and Cryptology}, pages 49--98. World Scientific, 2008.

\bibitem[Eisenbud(1995)]{bn:eisenbud}
D.~Eisenbud.
\newblock \emph{Commutative Algebra with a View Toward Algebraic Geometry},
  volume 150 of \emph{Graduate Texts in Mathematics}.
\newblock Springer-Verlag, Berlin, 1995.

\bibitem[Geil and Pellikaan(2002)]{geilpellikaan00}
O.~Geil and R.~Pellikaan.
\newblock On the structure of order domains.
\newblock \emph{Finite Fields Appl.}, 8\penalty0 (3):\penalty0 369--396, July
  2002.

\bibitem[Geil et~al.(2012)Geil, Matsumoto, and Ruano]{gmr12isit}
O.~Geil, R.~Matsumoto, and D.~Ruano.
\newblock List decoding algorithms based on Gr\"obner bases for general
  one-point AG codes.
\newblock In \emph{Proc. ISIT 2012}, pages 86--90, Cambridge, MA, USA, July
  2012.
\newblock arXiv:1201.6248.

\bibitem[Guruswami and Sudan(1999)]{guruswami99}
V.~Guruswami and M.~Sudan.
\newblock Improved decoding of Reed-Solomon and algebraic-geometry codes.
\newblock \emph{IEEE Trans. Inform. Theory}, 45\penalty0 (4):\penalty0
  1757--1767, Sept. 1999.

\bibitem[H\o holdt and Pellikaan(1995)]{hoholdt95}
T.~H\o holdt and R.~Pellikaan.
\newblock On the decoding of algebraic-geometric codes.
\newblock \emph{IEEE Trans. Inform. Theory}, 41\penalty0 (6):\penalty0
  1589--1614, Nov. 1995.

\bibitem[Lax(2012)]{lax12}
R.~F. Lax.
\newblock Generic interpolation polynomial for list decoding.
\newblock \emph{Finite Fields Appl.}, 18\penalty0 (1):\penalty0 167--178, Jan.
  2012.
\newblock \doi{10.1016/j.ffa.2011.07.007}.

\bibitem[Lee and O'Sullivan(2009)]{lee09}
K.~Lee and M.~E. O'Sullivan.
\newblock List decoding of Hermitian codes using Gr\"obner bases.
\newblock \emph{J. Symbolic Comput.}, 44\penalty0 (12):\penalty0 1662--1675,
  Dec. 2009.
\newblock arXiv:cs/0610132.

\bibitem[Little(2011)]{little11}
J.~B. Little.
\newblock List decoding for AG codes using Gr\"obner bases.
\newblock In \emph{SIAM Conference on Applied Algebraic Geometry}, North
  Carolina State University, NC, USA, Oct. 2011.

\bibitem[Matsumoto and Miura(2000{\natexlab{a}})]{matsumoto99ldpaper}
R.~Matsumoto and S.~Miura.
\newblock Finding a basis of a linear system with pairwise distinct discrete
  valuations on an algebraic curve.
\newblock \emph{J. Symbolic Comput.}, 30\penalty0 (3):\penalty0 309--323,
  Sept. 2000{\natexlab{a}}.

\bibitem[Matsumoto and Miura(2000{\natexlab{b}})]{miuraform}
R.~Matsumoto and S.~Miura.
\newblock On construction and generalization of algebraic geometry codes.
\newblock In T.~Katsura et~al., editors, \emph{Proc. Algebraic Geometry,
  Number Theory, Coding Theory, and Cryptography}, pages 3--15, Univ. Tokyo,
  Japan, Jan. 2000{\natexlab{b}}.
\newblock URL \url{http://www.rmatsumoto.org/repository/weight-construct.pdf}.

\bibitem[Miura(1993)]{miura92}
S.~Miura.
\newblock Algebraic geometric codes on certain plane curves.
\newblock \emph{Electronics and Communications in Japan (Part III: Fundamental
  Electronic Science)}, 76\penalty0 (12):\penalty0 1--13, Dec. 1993.
\newblock \doi{10.1002/ecjc.4430761201}.
\newblock (original Japanese version published as Trans. IEICE, vol. J75-A,
  no. 11, pp. 1735--1745, Nov. 1992).

\bibitem[Miura(1998)]{miura98}
S.~Miura.
\newblock Linear codes on affine algebraic curves.
\newblock \emph{Trans. IEICE}, J81-A\penalty0 (10):\penalty0 1398--1421, Oct.
  1998.

\bibitem[Rosales and Garc\'ia-S\'anchez(2009)]{MR2549780}
J.~C. Rosales and P.~A. Garc\'ia-S\'anchez.
\newblock \emph{Numerical Semigroups}, volume~20 of \emph{Developments in
  Mathematics}.
\newblock Springer, New York, 2009.
\newblock ISBN 978-1-4419-0159-0.

\bibitem[Saints and Heegard(1995)]{saints95}
K.~Saints and C.~Heegard.
\newblock Algebraic-geometric codes and multidimensional cyclic codes: A
  unified theory and algorithms for decoding using Gr\"obner bases.
\newblock \emph{IEEE Trans. Inform. Theory}, 41\penalty0 (6):\penalty0
  1733--1751, Nov. 1995.

\bibitem[Sakata(2001)]{sakata01}
S.~Sakata.
\newblock On fast interpolation method for {Guruswami-Sudan} list decoding of
  one-point algebraic-geometry codes.
\newblock In S.~Bozta\c{s} and I.~E. Shparlinski, editors, \emph{Proc.\
  AAECC-14}, volume 2227 of \emph{Lecture Notes in Computer Science}, pages
  172--181, Melbourne, Australia, Nov. 2001. Springer-Verlag.
\newblock \doi{10.1007/3-540-45624-4_18}.

\bibitem[Schicho(1998)]{schicho98}
J.~Schicho.
\newblock Inversion of birational maps with Gr\"obner bases.
\newblock In B.~Buchberger and F.~Winkler, editors, \emph{Gr\"obner Bases
  and Applications}, volume 251 of \emph{London Mathematical Society Lecture
  Note Series}, pages 495--503. Cambridge University Press, 1998.
\newblock ISBN 9780521632980.
\newblock \doi{10.1017/CBO9780511565847.031}.

\bibitem[Stichtenoth(1993)]{bn:stichtenoth}
H.~Stichtenoth.
\newblock \emph{Algebraic Function Fields and Codes}.
\newblock Springer-Verlag, Berlin, 1993.

\bibitem[Tang(1998)]{tang98}
L.-Z. Tang.
\newblock A Gr\"obner basis criterion for birational equivalence of affine
  varieties.
\newblock \emph{J. Pure Appl. Algebra}, 123\penalty0 (1-3):\penalty0
  275--283, Jan. 1998.
\newblock \doi{10.1016/S0022-4049(97)00139-4}.

\bibitem[Vasconcelos(1998)]{bn:vasconcelos}
W.~V. Vasconcelos.
\newblock \emph{Computational Methods in Commutative Algebra and Algebraic
  Geometry}, volume~2 of \emph{Algorithms and Computation in Mathematics}.
\newblock Springer-Verlag, Berlin, 1998.

\end{thebibliography}
\end{document}